\documentclass[aps,pra,floatfix,showpacs,12pt]{revtex4-1}

\usepackage{graphicx}
\usepackage{amsmath, amsfonts, amssymb, bm}

\usepackage{amstext}
\usepackage{mathrsfs}
\usepackage{color}

\begin{document}
\title{Analytical tools for investigating strong-field QED processes in tightly focused laser fields}
\author{A.~Di Piazza}
\email{dipiazza@mpi-hd.mpg.de}
\affiliation{Max-Planck-Institut f\"ur Kernphysik, Saupfercheckweg 1, D-69117 Heidelberg, Germany}

\begin{abstract}
The present paper is the natural continuation of the letter [Phys. Rev. Lett. \textbf{113}, 040402 (2014)], where the electron wave functions in the presence of a background electromagnetic field of general space-time structure have been constructed analytically, assuming that the initial energy of the electron is the largest dynamical energy scale in the problem and having in mind the case of a background tightly focused laser beam. Here, we determine the scalar and the spinor propagators under the same approximations, which are useful tools for calculating, e.g., total probabilities of processes occurring in such complex electromagnetic fields. In addition, we also present a simpler and more general expression of the electron wave functions found in [Phys. Rev. Lett. \textbf{113}, 040402 (2014)] and we indicate a substitution rule to obtain them starting from the well-known Volkov wave functions in a plane-wave field.
\end{abstract}

\pacs{12.20.Ds, 41.60.-m}
\maketitle

\section{Introduction}
Quantum Electrodynamics (QED) is an extremely successful theory and its predictions have been confirmed experimentally with outstanding precision. This success has corroborated the general approach originally developed by Feynman, Schwinger, and Tomonaga, and based on the renormalization theory (see, e.g., the monographs \cite{Jauch_b_1976,Landau_b_4_1982,Peskin_b_1995,Mandl_b_2010}). Feynman, Schwinger, and Tomonaga's approach was extended by Furry \cite{Furry_1951} to the description of processes occurring in the presence of some classes of classical, background electromagnetic fields (these classes exclude, for example, those electromagnetic fields, which render the QED vacuum unstable under electron-positron pair production, see \cite{Fradkin_b_1991}). The application of the so-called Furry picture is required when the interaction of electrons (and positrons) with the external field has to be taken into account exactly. The corresponding mathematical and physical conditions depend on the structure of the background field. Here, we are interested in processes occurring in the presence of a strong laser beam and the controlling parameter is the so-called classical nonlinearity parameter $\xi=|e|F_0/m\omega_0$, where $F_0$ is the amplitude of the laser wave and $\omega_0$ is its central angular frequency ($m$ and $e<0$ are the electron mass and charge, respectively, and units with $c=1$ are employed throughout) \cite{Mitter_1975,Ritus_1985,Ehlotzky_2009,Di_Piazza_2012}. If $\xi\gtrsim 1$, it is necessary to include the effects of the laser field exactly in the calculations from the beginning. Since a realistic laser field has a complex space-time structure, this is in general an impossible task as it requires the exact analytical solution of the Dirac equation in the presence of the background laser field. However, the task has been accomplished by Volkov in the ideal case of a laser field approximated by a plane wave \cite{Volkov_1935} (see also \cite{Landau_b_4_1982}). Starting from the seminal paper \cite{Reiss_1962}, the corresponding electron wave functions (Volkov states) have been extensively employed to investigate various strong-field QED processes in a plane-wave field from the first-order ones as nonlinear Compton scattering \cite{Nikishov_1964,Goldman_1964,Narozhny_2000,Boca_2009,Harvey_2009,Heinzl_2009,Mackenroth_2010,Mackenroth_2011,Seipt_2011,Krajewska_2012,Seipt_2013,Dinu_2013,Nedoreshta_2013,Krajewska_2014,Wistisen_2014}, nonlinear Breit-Wheeler pair production \cite{Nikishov_1964,Narozhny_2000,Roshchupkin_2001,Heinzl_2010b,Mueller_2011b,Titov_2012,Nousch_2012,Krajewska_2013b,Jansen_2013,Augustin_2014}, nonlinear Bethe-Heitler pair production \cite{Yakovlev_1966,MVG2003PRL,MVG2003PRA,Avetissian_2003,Kaminski_2006,Krajewska_2006,Muller2009PLB}, and electron-positron annihilation \cite{Ilderton_2011b}, to higher-order processes like nonlinear Double Compton scattering \cite{Loetstedt_2009,Seipt_2012,Mackenroth_2013} and trident pair production \cite{Hu_2010,Ilderton_2011}. In order to calculate a second-order process like nonlinear double Compton scattering, the Green's function of the Dirac equation in the presence of a plane wave (Volkov propagator) has to be used \cite{Brown_1964,Reiss_1966,Eberly_1966,Mitter_1975,Ritus_1985}. The Volkov propagator is also a useful tool to investigate radiative corrections in the presence of a laser field like the polarization operator \cite{Becker_1975} (see also \cite{Meuren_2013}). It is remarkable, however, that originally the polarization operator has been determined independently in \cite{Baier_1976_b} by employing the operator technique, which avoids the direct use of the Volkov propagator. The same operator technique has been employed in \cite{Baier_1976_a} to calculate the mass operator in a plane wave and in \cite{Di_Piazza_2007} to investigate the process of photon splitting in a plane wave. The representation of the mass operator and of the polarization operator found in \cite{Baier_1976_a} and in \cite{Baier_1976_b} via the operator technique has been 
later widely exploited to study radiative corrections and vacuum-polarization effects in plane-wave fields \cite{Di_Piazza_2008,Meuren_2011,Di_Piazza_2013,Dinu_2014,Dinu_2014b,Gies_2014} and the total rate of nonlinear Breit-Wheeler pair production via the optical theorem \cite{Milstein_2006} (see also \cite{Meuren_2015}).

We emphasize that nowadays the parameter region $\xi\gtrsim 1$, where the laser field has to be taken into account exactly in determining the electron dynamics, can be routinely entered experimentally by means of optical laser beams. In fact, the threshold $\xi=1$ at an optical laser photon energy of $1\;\text{eV}$ requires a laser intensity of the order of $10^{18}\;\text{W/cm$^2$}$ and the present record intensity achieved in the laboratory exceeds this value by about four orders of magnitude \cite{Yanovsky_2008}. On the other hand, we point out that the condition $\xi\gtrsim 1$ is not related to the importance of pure quantum effects like the photon recoil in the emission of radiation or the creation of an electron-positron pair in laser-photon collision. The importance of such effects is determined by the ratio of the laser field amplitude $F_0$ and the the so-called critical field of QED: $F_{cr}=m^2/\hbar|e|=1.3\times 10^{16}\;\text{V/cm}=4.4\times 10^{13}\;\text{G}$ \cite{Landau_b_4_1982,Baier_b_1998,Fradkin_b_1991}. The ratio $F_0/F_{cr}$ in the laboratory frame is not a Lorentz invariant quantity and therefore it cannot represent by itself the physical parameter controlling the importance of quantum effects. In the case of processes primed by an electron with initial four-momentum $p^{\mu}$ and colliding with a plane-wave field, it turns out that the important parameter is the so-called quantum nonlinearity parameter $\chi=((k_0p)/m\omega_0)(F_0/F_{cr})$ \cite{Mitter_1975,Ritus_1985,Ehlotzky_2009,Di_Piazza_2012}, where $\hbar k_0^{\mu}$ is the laser photons four-momentum. The parameter $\chi$ represents the amplitude of the laser field in units of the critical field of QED in the initial rest-frame of the electron. Analogously, in the case of nonlinear Breit-Wheeler pair production by a photon with four-momentum $q^{\mu}$, the quantum nonlinearity parameter is $\kappa=((k_0q)/m\omega_0)(F_0/F_{cr})$ \cite{Mitter_1975,Ritus_1985,Ehlotzky_2009,Di_Piazza_2012}, which corresponds to the amplitude of the plane-wave field in units of $F_{cr}$ in the final center-of-momentum frame of the created electron and positron. Entering the strong-field QED regime where not only $\xi\gtrsim 1$ but also $\chi\gtrsim 1$ (here, we limit to processes primed by an electron), is a challenging task nowadays as the amplitude of the critical field of QED largely exceeds available field strengths (the critical field amplitude corresponds to a laser intensity of the order of $10^{29}\;\text{W/cm$^2$}$). Thus, even considering upcoming 10-PW facilities as APOLLON-10P \cite{APOLLON_10P} (expected intensities of the order of $10^{23}\;\text{W/cm$^2$}$) or future lasers like the Extreme Light Infrastructure (ELI) \cite{ELI} and the Exawatt Center for Extreme Light Studies (XCELS) \cite{XCELS} (expected intensities of the order of $10^{24}\text{-}10^{26}\;\text{W/cm$^2$}$), this regime can be entered only by employing ultra-relativistic electrons. In this respect, it is also worth pointing out that the expected intensities are assumed to be achieved by spatially focusing the laser energy almost down to the diffraction limit. Unavoidably, also temporal compression of the laser energy into few-cycle pulses is required but Volkov states and propagators can still be applied for temporally compressed plane waves. On the contrary, they cannot be employed if the background field is a spatially focused laser beam. We mention here that the so-far unique experimental campaign on strong-laser field QED \cite{Bula_1996,Burke_1997} has been carried out with a laser of wavelength of the order of $1\;\text{$\mu$m}$ and a relatively large pulse spot area ($\sim 60\;\text{$\mu$m$^2$}$), such that the experimental results were well reproduced theoretically within the plane-wave field approximation.

The requirement of tightly focusing the laser energy also in space in order to reach experimentally ultra-high intensities has motivated us to investigate the possibility of determining analytically the electron wave functions for tightly focused laser fields under certain realistic approximations (as we have mentioned, the problem in general does not admit an exact analytical solution) \cite{Di_Piazza_2014}. In \cite{Di_Piazza_2014}, we have assumed that the electron is ultra-relativistic and that its initial energy $\varepsilon$ is the largest dynamical energy in the problem \footnote{A basic assumption of the approach followed in \cite{Di_Piazza_2014} and here is to neglect the instability of vacuum in the presence of the focused laser beam. Such instability starts playing a role at intensities of the order of $10^{27}\;\text{W/cm$^2$}$ \cite{Narozhny_2004}, which lie well beyond the intensities we want to consider.}. We emphasize that the formalism employed in \cite{Di_Piazza_2014} and here below does not apply only to the electron but rather to any charged particle, provided that the dynamical conditions of validity of the approach are fulfilled (implicitly one also requires, of course, that the particle can be assumed to be stable for the whole duration of the interaction). In this respect, although below we will refer for definiteness to the electron and the wave functions will contain the electron charge and mass, it should be borne in mind that the results can also be employed in principle for other charged particles. Now, the meaning of the assumption on the initial energy of the electron being the largest dynamical energy in the problem was to ensure that the electron is barely deviated from its initial direction, which was chosen as the reference direction for introducing light-cone coordinates. In the experimentally relevant case of an electron initially counterpropagating with respect to a tightly focused laser beam, the assumption implied in particular that although the laser amplitude is such that $\xi\gg 1$, as it is envisaged in planned experiment on strong-field QED, the transverse momentum of the electron in the laser field (of the order of $m\xi$) still remains much smaller than $\varepsilon$. Under this approximation, we constructed the electron wave functions within the Wentzel-Kramers-Brillouin (WKB) approximation, by including terms scaling as $1/\varepsilon$ \cite{Landau_b_2_1975,Landau_b_3_1977}. We notice that the above assumption $\varepsilon\gg m\xi$ well fits present and near-future experimental conditions envisaged to test strong-field QED with intense lasers. For example, even at the next-generation facility APOLLON-10P the quantity $m\xi$ will be of the order of $80\;\text{MeV}$ \cite{APOLLON_10P} and, in order to reach the threshold $\chi=1$ where nonlinear QED effects become essential, electron energies of about $500\;\text{MeV}$ are required. Moreover, electron beams with energies of about $4\;\text{GeV}$ have been recently demonstrated experimentally with laser-plasma accelerators \cite{Leemans_2014} (conventional accelerators provide electron beams with energies even exceeding $200\;\text{GeV}$, see, e.g., \cite{Uggerhoj_2005}). It is worth pointing out that the results in \cite{Di_Piazza_2014} and those presented here are also valid if $\xi\sim 1$. Now, at $\xi\sim 1$ the formation region of strong-field QED processes as nonlinear Compton scattering and nonlinear Breit-Wheeler pair production in a plane-wave field is of the order of the laser wavelength \cite{Ritus_1985} and the tight focusing of the laser field is expected to essentially alter the probabilities with respect to the plane-wave results. On the contrary, in the regime $\xi\gg 1$ the formation region of the mentioned strong-field QED processes in a plane wave has been found to be $\xi$ times smaller than the laser wavelength \cite{Ritus_1985}. In this regime, one would expect, for example, that the total probability of the mentioned strong-field QED processes in a tightly focused laser pulse can be obtained by averaging the probability calculated in the presence of a constant crossed field, after replacing the latter with the local value of the tightly focused laser field (local constant-crossed field approximation) \cite{Ritus_1985}. However, it has been found recently that the local constant-crossed field approximation in some circumstances does not reproduce 
accurately the harmonic structures of radiation emission spectra in nonlinear Compton scattering \cite{Harvey_2015}. Although the differences might be explained as a consequence of interference effects among different emission points \cite{Ritus_1985,Meuren_2015b}, such interference effects are non-local and this also calls for a thorough analytical investigation of strong-field QED processes in the presence of a tightly focused laser beam. Moreover, such an analysis can also shed light on the accuracy of the local constant-crossed field approximation at intermediate values of the parameter $\xi$ and for those fields where the local value of the parameter $\xi$ varies from large values to values of the order of or less than unity.

In the present paper, we continue and extend the investigation started in \cite{Di_Piazza_2014} in the following directions. First, in Section \ref{WF} we present a simpler expression of the wave functions found in \cite{Di_Piazza_2014}, which is equivalent under the approximations employed. Moreover, we generalize the previous results allowing for a longitudinal direction of light-cone coordinates non necessarily coinciding with the initial momentum of the electron. Second, in Section \ref{Prop}, we construct the corresponding scalar and spinor propagator under the same approximations. We note that a general expression of the WKB propagator for a scalar particle has been obtained in \cite{Yakaboylu_2014}, by applying the so-called path-dependent formulation of a gauge theory based on the path-integral representation of the propagator. Here, we employ the operator technique, which we have already applied, e.g., in \cite{Di_Piazza_2012b,Di_Piazza_2014b} to calculate the electron quasiclassical wave functions and propagator in the presence of a combined atomic and plane-wave field. After determining the propagator, we show that the found wave functions can be obtained by applying a limiting procedure to the propagator itself. Finally, in Section \ref{Concl}, the main conclusions of the paper and possible future investigations are presented.

%
%
\section{Ultra-relativistic WKB electron wave functions in a focused laser field}
\label{WF}
We consider an ultrarelativistic electron moving in a background electromagnetic field, described by the four-vector potential $A^{\mu}(x)$ in the Lorentz gauge $\partial_{\mu}A^{\mu}=0$. The four-vector $p^{\mu}=(\varepsilon,\bm{p})$, with $\varepsilon=\sqrt{m^2+\bm{p}^2}$, indicates the asymptotic four-momentum of the electron, i.e., either in the remote past or in the remote future. Even though the considerations below apply to electromagnetic fields of rather general space-time structure, we have in mind the case where the background electromagnetic field represents an intense, few-cycle, and tightly focused laser beam. Thus, we also assume that the field tensor $F^{\mu\nu}(x)=\partial^{\mu}A^{\nu}(x)-\partial^{\nu}A^{\mu}(x)=(\bm{E}(x),\bm{B}(x))$ is localized on a space and a time region of a few cubic typical laser wavelengths and a few typical laser periods, respectively, and that it has a maximum amplitude $F_0$. The typical laser wavelength and period correspond to a typical laser angular frequency, which is indicated as $\omega_0$ and which is such that the classical nonlinearity parameter $\xi$ satisfies the strong inequalities: $m\ll m\xi\ll\varepsilon$ (see \cite{Di_Piazza_2014}). As we have already hinted in the Introduction, under the above assumptions the electron is only slightly deflected from its asymptotic direction by the background field in the physically relevant situation where it is initially counterpropagating with respect to the laser field and, from this point of view, the regime $\xi\sim 1$ can also be investigated with the present method. Thus, it is convenient to introduce the light-cone coordinates 
\begin{align}
T=\frac{t+\bm{n}\cdot \bm{x}}{2}, && \bm{x}_{\perp}=\bm{x}-(\bm{n}\cdot \bm{x})\bm{n}, && \phi=t-\bm{n}\cdot \bm{x}
\end{align}
for a space-time point $x$ with space-time coordinates $x^{\mu}=(t,\bm{x})$ and to indicate the four light-cone coordinates via the corresponding capital letter $X$: $X^{\mu}=(T,\bm{x}_{\perp},\phi)$. In \cite{Di_Piazza_2014}, for the sake of simplicity, we have chosen $\bm{n}$ to coincide with $\bm{p}/|\bm{p}|$ as this essentially simplifies the solution of the classical equations of motion. However, as it will be clear below, this assumption is not strictly necessary and more general expressions of the electron wave functions can be obtained by assuming that the direction $\bm{n}$ is such that $|\bm{p}_{\perp}|\lesssim m\xi\ll p_+$, where $\bm{p}_{\perp}=\bm{p}-(\bm{n}\cdot\bm{p})\bm{n}$ and $p_+=(\varepsilon+\bm{n}\cdot\bm{p})/2$ are three of the four light-cone coordinates of the asymptotic four-momentum. The on-shell condition $p^2=m^2$ implies that the remaining light-cone coordinate $p_-=\varepsilon-\bm{n}\cdot\bm{p}$ can be expressed in terms of the other three as $p_-=(m^2+\bm{p}_{\perp}^2)/2p_+$. The light-cone coordinates of an arbitrary four-vector $v^{\mu}=(v^0,\bm{v})$ are defined in the same way as for the four-momentum: $v_+=(v^0+\bm{n}\cdot\bm{v})/2$, $\bm{v}_{\perp}=\bm{v}-(\bm{n}\cdot \bm{v})\bm{n}$, and $v_-=v^0-\bm{n}\cdot\bm{v}$. The same definition is extended to the Dirac gamma matrices $\gamma^{\mu}=(\gamma^0,\bm{\gamma})$, which are intended in the Dirac representation: $\gamma_+=(\gamma^0+\bm{n}\cdot\bm{\gamma})/2$, $\bm{\gamma}_{\perp}=\bm{\gamma}-(\bm{n}\cdot\bm{\gamma})\bm{n}$, and $\gamma_-=\gamma^0-\bm{n}\cdot\bm{\gamma}$. Moreover, the derivatives 
\begin{align}
\frac{\partial}{\partial T}=\frac{\partial}{\partial t}+\frac{\partial}{\partial (\bm{n}\cdot \bm{x})}, && \frac{\partial}{\partial \phi}=\frac{1}{2}\left(\frac{\partial}{\partial t}-\frac{\partial}{\partial (\bm{n}\cdot \bm{x})}\right)
\end{align}
with respect to the light-cone coordinates $T$ and $\phi$ can be derived from the relations
\begin{align}
\frac{\partial}{\partial t}=\frac{1}{2}\frac{\partial}{\partial T}+\frac{\partial}{\partial \phi},&& \frac{\partial}{\partial (\bm{n}\cdot \bm{x})}=\frac{1}{2}\frac{\partial}{\partial T}-\frac{\partial}{\partial \phi},
\end{align}
whereas the derivatives with respect to the transverse coordinates form the two-dimensional vector $\bm{\nabla}_{\perp}=\partial/\partial \bm{x}_{\perp}$. Also, the scalar product between two four-vectors $u^{\mu}$ and $v^{\mu}$ can be written as 
\begin{equation}
(uv)=u_+v_-+u_-v_+-\bm{u}_{\perp}\cdot\bm{v}_{\perp}
\end{equation}
and the four-divergence of a vector field $V^{\mu}(x)$ as
\begin{equation}
\partial_{\mu} V^{\mu}=\frac{\partial V_+}{\partial T}+\frac{\partial V_-}{\partial \phi}+\bm{\nabla}_{\perp}\cdot\bm{V}_{\perp}.
\end{equation}

Below, for the sake of completeness, we summarize the important steps we followed in \cite{Di_Piazza_2014} in order to solve the Dirac equation
\begin{equation}
\label{Dirac}
[\gamma^{\mu}(i\hbar\partial_{\mu}-eA_{\mu})-m]\psi=0,
\end{equation}
for specific asymptotic conditions on the electron wave function $\psi(x|A)$ and within the parameter regime specified above. Recall that in terms of the light-cone coordinates the largest dynamical energy scale of the problem is $p_+$, which is approximately equal to $\varepsilon$ in the ultra-relativistic regime. Now, based on the general argument that the de Broglie wavelength of an ultra-relativistic particle is very small, we applied the WKB method \cite{Landau_b_3_1977} and we looked for a solution of the Dirac equation of the form $\psi(x|A)=\exp[iS(x|A)/\hbar]\varphi(x|A)$ \cite{Pauli_1932,Rubinow_1963,Maslov_b_1981}. In \cite{Di_Piazza_2014}, we have observed that, under the assumption that $m\ll m\xi\ll \varepsilon$, the WKB method can be applied if the condition $\hbar/w_0\varepsilon\ll 1$ is fulfilled, with $w_0$ being the laser waist size. This condition is safely satisfied in the case of a tightly focused, optical ($w_0\sim 1\;\text{$\mu$m}$) laser field, as $\hbar/\varepsilon\sim m\lambda_C/\varepsilon\ll\lambda_C$, with $\lambda_C=\hbar/m=3.9\times 10^{-11}\;\text{cm}$ being the Compton wavelength. By writing the electron wave function as $\psi(x|A)=\exp[iS(x|A)/\hbar]\varphi(x|A)$,  the Dirac equation becomes
\begin{equation}
[\gamma^{\mu}(\partial_{\mu}S+eA_{\mu})+m]\varphi=i\hbar\gamma^{\mu}\partial_{\mu}\varphi.
\end{equation}
By first neglecting the term proportional to $\hbar$, one obtains
\begin{equation}
[\gamma^{\mu}(\partial_{\mu}S+eA_{\mu})+m]\varphi=0
\end{equation}
for the bispinor $\varphi(x|A)$ at zero-order in $\hbar$. This equation admits a non-vanishing solution only if $\det[\gamma^{\mu}(\partial_{\mu}S+eA_{\mu})+m]=0$, which implies that the quantity $S(x|A)$ has to satisfy the Hamilton-Jacobi equation
\begin{equation}
(\partial_{\mu}S+eA_{\mu})(\partial^{\mu}S+eA^{\mu})-m^2=0
\end{equation}
and that it can be identified with the classical action \cite{Landau_b_2_1975}. In \cite{Di_Piazza_2014}, starting from the solution of the classical equations of motion, we employed the method of characteristics in order to construct the action $S(x|A)$. Here, we follow a different procedure, which allows us to easily obtain a more general expression of the action, including the possibility that $\bm{p}_{\perp}$ does not vanish as it was in \cite{Di_Piazza_2014}. First, we pass to light-cone coordinates and we initially assume that $p^{\mu}$ is the electron four-momentum at a fixed reference ``time'' $T_0$, which will be set to $\pm \infty$ later. Thus, we write the action as $S(X|A)=S_p(X;T_0|A)=-(p_+\phi+p_-T-\bm{p}_{\perp}\cdot\bm{x}_{\perp})+\delta S_p(X;T_0|A)$. The quantity $\delta S_p(X;T_0|A)$ satisfies the equation
\begin{equation}
\label{HJ_lc}
\begin{split}
p_+&\left(\frac{\partial\delta S_p}{\partial T}+eA_-\right)+\bm{p}_{\perp}\cdot(\bm{\nabla}_{\perp}\delta S_p-e\bm{A}_{\perp})-\frac{\partial\delta S_p}{\partial \phi}\frac{\partial\delta S_p}{\partial T}+\frac{1}{2}(\bm{\nabla}_{\perp}\delta S_p)^2-e^2A_-A_++\frac{1}{2}e^2\bm{A}_{\perp}^2\\
&-eA_-\frac{\partial\delta S_p}{\partial \phi}-eA_+\frac{\partial\delta S_p}{\partial T}-e\bm{A}_{\perp}\cdot\bm{\nabla}_{\perp}\delta S_p+p_-\left(\frac{\partial\delta S_p}{\partial \phi}+eA_+\right)=0.
\end{split}
\end{equation}
This equation can now be solved iteratively by setting $\delta S_p(X;T_0|A)=\delta S_p^{(0)}(X;T_0|A)+\delta S_p^{(1)}(X;T_0|A)+\cdots$, with $\delta S_p^{(n)}(X;T_0|A)=O(1/p_+^n)$. It is clear that at the leading order we can write
\begin{equation}
\delta S_p^{(0)}(X;T_0|A)=-e\int_{T_0}^TdT' A_-(T',\bm{x}_{\perp},\phi),
\end{equation}
where the same initial conditions as in \cite{Di_Piazza_2014} have been assumed for the four-vector potential, i.e., $A^{\mu}(T_0,\bm{x}_{\perp},\phi)=0$. By substituting this expression of $\delta S_p^{(0)}(X;T_0|A)$ into Eq. (\ref{HJ_lc}), it is easy to show that the first-order correction $\delta S_p^{(1)}(X;T_0|A)$ reads
\begin{equation}
\delta S_p^{(1)}(X;T_0|A)=\frac{1}{p_+}\int_{T_0}^TdT' \left[e\bm{p}_{\perp}\cdot\bm{\mathcal{A}}_{\perp}(T',\bm{x}_{\perp},\phi;T_0)-
\frac{1}{2}e^2\bm{\mathcal{A}}_{\perp}^2(T',\bm{x}_{\perp},\phi;T_0)\right],
\end{equation}
with
\begin{equation}
\bm{\mathcal{A}}_{\perp}(X;T_0)=\bm{A}_{\perp}(X)+\bm{\nabla}_{\perp}\int_{T_0}^TdT'A_-(T',\bm{x}_{\perp},\phi).
\end{equation}
In order to relate the results below to the finding in \cite{Di_Piazza_2014}, we observe that the quantity $\bm{\mathcal{A}}_{\perp}(X;T_0)$ can be written in terms of the electromagnetic field as
\begin{equation}
\bm{\mathcal{A}}_{\perp}(X;T_0)=-\int_{T_0}^TdT'[\bm{E}_{\perp}(T',\bm{x}_{\perp},\phi)+\bm{n}\times\bm{B}_{\perp}(T',\bm{x}_{\perp},\phi)],
\end{equation}
which is equal to $-\bm{G}_{p}(\tau,\bm{r}_0)$ in the notation employed in \cite{Di_Piazza_2014}. Also, we notice that up to first order in $1/p_+$, the dependence on $\phi$ can be neglected in the external four-potential. In fact, as it can be seen in Eq. (3) in \cite{Di_Piazza_2014}, it is $\phi(T)-\phi_0=O(1/p_+^2)$. Correspondingly, Eq. (6) there shows that, by neglecting the derivatives of the four-potential with respect to $\phi$, the conjugated momentum $p_+-eA_+(T,\bm{x}_{\perp})$ is conserved up to terms $O(1/p_+)$ with respect to the leading one. Since here we will neglect the dependence on $\phi$ of the four-vector potential (one can assume for simplicity that $\phi_0=0$), it is convenient to introduce the three-dimensional light-cone coordinates $\bm{X}=(T,\bm{x}_{\perp})$ and define the effective transverse vector potential
\begin{equation}
\bm{\mathcal{A}}_{\perp}(\bm{X};T_0)=\bm{A}_{\perp}(\bm{X})+\bm{\nabla}_{\perp}\int_{T_0}^TdT'A_-(T',\bm{x}_{\perp}).
\end{equation}
Thus, we can write the action $S_p(X;T_0|A)$ up to first order terms in $1/p_+$ as (for the sake of notational simplicity and since no confusion can arise, from now on we do not write explicitly the upper index $(1)$)
\begin{equation}
\label{S}
\begin{split}
S_p(X;T_0|A)=&-(p_+\phi+p_-T-\bm{p}_{\perp}\cdot\bm{x}_{\perp})-e\int_{T_0}^TdT' A_-(T',\bm{x}_{\perp})\\
&+\frac{1}{p_+}\int_{T_0}^TdT' \left[e\bm{p}_{\perp}\cdot\bm{\mathcal{A}}_{\perp}(T',\bm{x}_{\perp};T_0)-
\frac{1}{2}e^2\bm{\mathcal{A}}_{\perp}^2(T',\bm{x}_{\perp};T_0)\right],
\end{split}
\end{equation}
which generalizes the expression found in \cite{Di_Piazza_2014}. It is also worth observing that the present approach allows to easily find higher-order corrections in $1/p_+$ in the action by means of Eq. (\ref{HJ_lc}).

After determining the action, the construction  of the first-order electron wave function $\psi_{p,\sigma}(X|A)=\psi_{p,\sigma}(X;T_0|A)$, with $\sigma$ being the spin quantum number at $T_0$, proceeds as in \cite{Di_Piazza_2014} and the final result is
\begin{equation}
\label{psi_0}
\begin{split}
\psi_{p,\sigma}(X;T_0|A)=&e^{iS_p(X;T_0|A)/\hbar}\bm{\bigg(}1+\frac{e}{2}\int_{T_0}^T\frac{dT'}{p_+}\{\bm{\nabla}_{\perp}\cdot\bm{\mathcal{A}}_{\perp}(T',\bm{x}_{\perp};T_0)\\
&+i\bm{\Sigma}\cdot[\bm{B}(T',\bm{x}_{\perp})-\bm{n}\times\bm{E}(T',\bm{x}_{\perp})]\}\bm{\bigg)} \frac{u_{p,\sigma}}{\sqrt{2\varepsilon}},
\end{split}
\end{equation}
where $\bm{\Sigma}=-i\gamma^1\gamma^2\gamma^3\bm{\gamma}$, $u_{p,\sigma}$ is the usual constant free bispinor normalized  as $u^{\dag}_{p,\sigma}u_{p,\sigma}=2\varepsilon$ \cite{Landau_b_4_1982}, and a unit quantization volume has been assumed. Apart from the more general expression of the action $S_p(X;T_0|A)$, this expression of the wave function is equivalent to the one found in \cite{Di_Piazza_2014}. 

Now, as it is observed, e.g., in \cite{Landau_b_4_1982}, when performing the matrix elements of basic strong-field QED processes like nonlinear Compton scattering and nonlinear Breit-Wheeler pair production in the ultra-relativistic regime, the contribution of those terms in the pre-exponent proportional to the unit matrix is suppressed with respect to those having a different matrix structure. The reason is that, since all particles involved in such processes almost propagate along the same direction and thus the photon polarization vectors $\bm{e}_{\lambda}$, with $\lambda=1,2$, almost lie on the plane perpendicular to this direction, the matrix element of the interaction matrix $\bm{\gamma}\cdot\bm{e}_{\lambda}$ between two constant bispinors vanishes at leading order in $1/p_+$. As a result, the leading-order terms proportional to the unit matrix and the terms $O(1/p_+)$ with a different matrix structure in the pre-exponent in Eq. (\ref{psi_0}) give contributions of the same order of magnitude to the transition matrix elements. Thus, in order to calculate probabilities of strong-field QED processes at the leading order within our approximations, all first-order corrections in the pre-exponent proportional to the unit matrix can be neglected. This is also formally corroborated by observing that, from the derivation presented in \cite{Di_Piazza_2014}, the pre-exponential function in Eq. (\ref{psi_0}) proportional to the unit matrix results from the overall exponential factor $\exp\Big[(1/2p_+)\int_{T_0}^TdT'\partial_{\mu}\partial^{\mu}S_p(T',\bm{x}_{\perp},\phi;T_0|A)\Big]$ (see Eq. (11) in the Supplemental Material in \cite{Di_Piazza_2014}), which is proportional to the well-known Van Vleck determinant \cite{Van_Vleck_1928} (see also \cite{Schiller_1962a,Schiller_1962b}) and which can be neglected within our approximations, as corresponding to a quantum correction to the action. The above argumentation allows us to write the wave function $\psi_{p,\sigma}(X;T_0|A)$ in a simpler form than in Eq. (\ref{psi_0}) but equivalent within our approximations. In fact, by exploiting the approximated relation $\gamma^0u_{p,\sigma}\approx \bm{n}\cdot\bm{\gamma}u_{p,\sigma}$ in the first-order term in the pre-exponent in Eq. (\ref{psi_0}), it can be easily shown that the wave function $\psi_{p,\sigma}(X;T_0|A)$ can also be written as
\begin{equation}
\label{psi_0_f}
\psi_{p,\sigma}(X;T_0|A)=E_p(X;T_0|A)\frac{u_{p,\sigma}}{\sqrt{2\varepsilon}},
\end{equation}
where the matrix
\begin{equation}
\label{E_p}
E_p(X;T_0|A)=e^{iS_p(X;T_0|A)/\hbar}\bigg[1-\frac{e}{2p_+}\gamma_+\bm{\gamma}_{\perp}\cdot\bm{\mathcal{A}}_{\perp}(\bm{X};T_0)\bigg]
\end{equation}
is the analogous of the Ritus matrix in the case of a focused laser field (see \cite{Ritus_1985}).

This form of the wave function is very suggestive. In fact, apart from the inconsequential phase factor $\exp\Big[-i(e/\hbar)\int_{T_0}^TdT' A_-(T',\bm{x}_{\perp})\Big]$ (see also Eq. (\ref{S})), which does not contribute when evaluating transition probabilities of strong-field QED processes, the structure of the state $\psi_{p,\sigma}(X;T_0|A)$ is the same as that of a Volkov wave function for a plane wave propagating along the direction $-\bm{n}$ , and described by the four-vector potential $A^{\mu}(T)=(0,\bm{A}_{\perp}(T),0)$, with the substitution
\begin{equation}
\bm{A}_{\perp}(T)\to\bm{\mathcal{A}}_{\perp}(\bm{X};T_0).
\end{equation}
This turns out to be especially interesting, by calculating the local value $\chi(X)$ of the quantum nonlinearity parameter $\chi$ starting from its general definition \cite{Ritus_1985}
\begin{equation}
\chi(X)=\frac{1}{m F_{cr}}\sqrt{[\varepsilon\bm{E}(X)+\bm{p}\times\bm{B}(X)]^2-[\bm{p}\cdot\bm{E}(X)]^2}.
\end{equation}
In fact, up to the leading order in $1/p_+$, one obtains that $\chi(X)$ is independent of $\phi$ and it is given by
\begin{equation}
\chi(\bm{X})=\frac{p_+}{m}\frac{1}{F_{cr}}\left|\frac{\partial\bm{\mathcal{A}}_{\perp}(\bm{X};T_0)}{\partial T}\right|=\frac{p_+}{m}\frac{1}{F_{cr}}\left|\frac{\partial\bm{A}_{\perp}(\bm{X})}{\partial T}+\bm{\nabla}_{\perp}A_-(\bm{X})\right|,
\end{equation}
which exactly corresponds to the local value of the parameter $\chi$ for a ``plane wave'' with four-vector potential $A^{\mu}(\bm{X};T_0)=(0,\bm{\mathcal{A}}_{\perp}(\bm{X};T_0),0)$ (note that the laser phase corresponds to $2T$). This is in agreement with the general observation that an ultra-relativistic charged particle ``sees'' an arbitrary electromagnetic field locally as a plane wave \cite{Landau_b_2_1975}. Of course, the same correspondence also occurs by introducing a local definition $\xi(X)$ of the parameter $\xi$ as the ratio between $\chi(X)$ and the quantity $\hbar(kp)/m^2$ (see, e.g., \cite{Di_Piazza_2007}) or by starting from the manifestly Lorentz and gauge invariant definition of the parameter $\xi$ proposed in \cite{Heinzl_2009}. A related observation is in order here. In fact, in general, a Dirac bispinor $\psi(x|A)$, which is solution of the Dirac equation (\ref{Dirac}), transforms as $\psi(x|A)\to\psi'(x|A')=\exp(-ie g(x))\psi(x|A)$ under the gauge transformation $A^{\mu}(x)\to A^{\prime\mu}(x)=A^{\mu}(x)+\partial^{\mu}g(x)$. Now, let us consider the gauge transformation, which makes the component $A'_-(X)$ of the transformed four-vector potential vanish, such that the latter resembles the four-vector potential in the plane-wave case (recall that the component $A_+(X)$ does not play any role under our approximations). The corresponding gauge function can be chosen as $g(X;T_0)=-\int_{T_0}^TdT'A_-(T',\bm{x}_{\perp})$ and the transformed wave function resulting from the one in Eq. (\ref{psi_0_f}) would exactly correspond to a Volkov wave function in the plane-wave potential $\bm{A}_{\perp}(T)$ with the substitution $\bm{A}_{\perp}(T)\to\bm{A}'_{\perp}(\bm{X};T_0)(=\bm{\mathcal{A}}_{\perp}(\bm{X};T_0)$).

For the sake of completeness, we conclude this Section by writing the in- and out-states with positive and negative energy, which can be obtained from the general expressions in Eqs. (\ref{S}) and (\ref{psi_0_f}) by means of the limiting procedures $T_0\to\pm\infty$, as mentioned in \cite{Di_Piazza_2014}. By employing the same notation as in \cite{Landau_b_4_1982} for the negative-energy constant bispinors $u_{-p,-\sigma}$ normalized as $u^{\dag}_{-p,-\sigma}u_{-p,-\sigma}=2\varepsilon$, the final result is:
\begin{align}
\label{in-out_pm_i}
\psi^{(\text{in})}_{p,\sigma}(X|A)&=e^{iS^{(\text{in})}_p(X|A)/\hbar}\bigg[1-\frac{e}{2p_+}\gamma_+\bm{\gamma}_{\perp}\cdot\bm{\mathcal{A}}^{(\text{in})}_{\perp}(\bm{X})\bigg] \frac{u_{p,\sigma}}{\sqrt{2\varepsilon}},\\
\psi^{(\text{out})}_{p,\sigma}(X|A)&=e^{iS^{(\text{out})}_p(X|A)/\hbar}\bigg[1-\frac{e}{2p_+}\gamma_+\bm{\gamma}_{\perp}\cdot\bm{\mathcal{A}}^{(\text{out})}_{\perp}(\bm{X})\bigg] \frac{u_{p,\sigma}}{\sqrt{2\varepsilon}},\\
\psi^{(\text{in})}_{-p,-\sigma}(X|A)&=e^{iS^{(\text{in})}_{-p}(X|A)/\hbar}\bigg[1+\frac{e}{2p_+}\gamma_+\bm{\gamma}_{\perp}\cdot\bm{\mathcal{A}}^{(\text{in})}_{\perp}(\bm{X})\bigg] \frac{u_{-p,-\sigma}}{\sqrt{2\varepsilon}},\\
\label{in-out_pm_f}
\psi^{(\text{out})}_{-p,-\sigma}(X|A)&=e^{iS^{(\text{out})}_{-p}(X|A)/\hbar}\bigg[1+\frac{e}{2p_+}\gamma_+\bm{\gamma}_{\perp}\cdot\bm{\mathcal{A}}^{(\text{out})}_{\perp}(\bm{X})\bigg] \frac{u_{-p,-\sigma}}{\sqrt{2\varepsilon}},
\end{align}
where
\begin{align}
\begin{split}
S^{(\text{in})}_{\pm p}(X|A)&=\mp(p_+\phi+p_-T-\bm{p}_{\perp}\cdot\bm{x}_{\perp})-e\int_{-\infty}^TdT' A_-(T',\bm{x}_{\perp})\\
&+\frac{1}{p_+}\int_{-\infty}^TdT' \left[e\bm{p}_{\perp}\cdot\bm{\mathcal{A}}^{(\text{in})}_{\perp}(T',\bm{x}_{\perp})\mp
\frac{1}{2}e^2\bm{\mathcal{A}}^{(\text{in}),2}_{\perp}(T',\bm{x}_{\perp})\right],
\end{split}\\
\begin{split}
S^{(\text{out})}_{\pm p}(X|A)&=\mp(p_+\phi+p_-T-\bm{p}_{\perp}\cdot\bm{x}_{\perp})+e\int_T^{\infty}dT' A_-(T',\bm{x}_{\perp})\\
&-\frac{1}{p_+}\int_T^{\infty}dT' \left[e\bm{p}_{\perp}\cdot\bm{\mathcal{A}}^{(\text{out})}_{\perp}(T',\bm{x}_{\perp})\mp
\frac{1}{2}e^2\bm{\mathcal{A}}^{(\text{out}),2}_{\perp}(T',\bm{x}_{\perp})\right],
\end{split}\\
\bm{\mathcal{A}}^{(\text{in})}_{\perp}(\bm{X})&=\bm{A}_{\perp}(\bm{X})+\bm{\nabla}_{\perp}\int_{-\infty}^TdT'A_-(T',\bm{x}_{\perp}),\\
\bm{\mathcal{A}}^{(\text{out})}_{\perp}(\bm{X})&=\bm{A}_{\perp}(\bm{X})-\bm{\nabla}_{\perp}\int_T^{\infty}dT'A_-(T',\bm{x}_{\perp}).
\end{align}
%
%
%
\section{Ultra-relativistic WKB propagator in a focused laser field}
\label{Prop}
In the present section we derive the electron propagator under the same assumptions of Section \ref{WF} and we closely follow the approach developed in \cite{Di_Piazza_2012b,Di_Piazza_2014b}, based on the operator technique. The electron propagator $G(x,x'|A)$ satisfies the equation
\begin{equation}
[\gamma^{\mu}(i\hbar\partial_{\mu}-eA_{\mu}(x))-m]G(x,x'|A)=\delta(x-x').
\end{equation}
It is convenient to write the electron propagator in the form
\begin{equation}
\label{G_D}
G(x,x'|A)=[\gamma^{\mu}(i\hbar\partial_{\mu}-eA_{\mu}(x))+m]D(x,x'|A)
\end{equation}
such that the ``square'' propagator $D(x,x'|A)$ satisfies the equation
\begin{equation}
\label{D}
\left[(i\hbar\partial_{\mu}-eA_{\mu}(x))(i\hbar\partial^{\mu}-eA^{\mu}(x))-\frac{i}{2}e\hbar\sigma_{\mu\nu}F^{\mu\nu}(x)-m^2\right]D(x,x'|A)=\delta(x-x'),
\end{equation}
where $\sigma^{\mu\nu}=(1/2)[\gamma^{\mu},\gamma^{\nu}]$. Before carrying out the spinor case, one takes advantage in considering the simpler scalar case. By indicating as $D^{(0)}(x,x'|A)$ the corresponding propagator, one has to solve the equation
\begin{equation}
\label{D^0}
\left[(i\hbar\partial_{\mu}-eA_{\mu}(x))(i\hbar\partial^{\mu}-eA^{\mu}(x))-m^2\right]D^{(0)}(x,x'|A)=\delta(x-x').
\end{equation}
In order to work within the operator-technique approach, we introduce the momenta operators corresponding to the light-cone coordinates as
\begin{align}
P_{\phi}=-i\hbar\frac{\partial}{\partial \phi},&&
P_T=-i\hbar\frac{\partial}{\partial T}, && \bm{P}_{\perp}=-i\hbar\bm{\nabla}_{\perp},
\end{align}
and we express the propagator via the Schwinger proper-time representation
\begin{equation}
D^{(0)}(X,X'|A)=-\frac{i}{\hbar}\int_0^{\infty}dse^{i(s/\hbar)[2P_{\phi}P_T-\bm{P}^2_{\perp}+2e(A_+P_T+A_-P_{\phi}+\bm{A}_{\perp}\cdot\bm{P}_{\perp})+e^2A^2-m^2]}\delta(X-X').
\end{equation}
Here, the four-dimensional $\delta$-function $\delta(X-X')$ can be expressed as
\begin{equation}
\delta(X-X')=\int\frac{d^4p}{(2\pi\hbar)^4}e^{-i[p_+(\phi-\phi')+p_-(T-T')-\bm{p}_{\perp}\cdot(\bm{x}_{\perp}-\bm{x}'_{\perp})]/\hbar}
\end{equation}
and the four-vector $p^{\mu}$ can be identified with the four-momentum flowing through the propagator, which also explains the use of the symbol $p^{\mu}$ in accordance to the results in Section \ref{WF}. By replacing this expression of the $\delta$-function into the propagator, we obtain
\begin{equation}
\label{D^0_i}
\begin{split}
D^{(0)}(X,X'|A)=&-\frac{i}{\hbar}\int_0^{\infty}ds\int\frac{d^4p}{(2\pi\hbar)^4}e^{-i[p_+(\phi-\phi')+p_-(T-T')-\bm{p}_{\perp}\cdot(\bm{x}_{\perp}-\bm{x}'_{\perp})]/\hbar}e^{i(p^2-m^2)s/\hbar}\\
&\times \exp\left\{i\frac{s}{\hbar}[2P_{\phi}P_T-\bm{P}^2_{\perp}-2(p_+P_T+p_-P_{\phi}+\bm{p}_{\perp}\cdot\bm{P}_{\perp})\right.\\
&\qquad\qquad+2e[A_+(P_T-p_-)+A_-(P_{\phi}-p_+)+\bm{A}_{\perp}\cdot(\bm{P}_{\perp}+\bm{p}_{\perp})]+e^2A^2\Big\}.
\end{split}
\end{equation}
Analogously to what we have already carried out in the Section \ref{WF}, we will apply the propagator to the investigation of processes, where the particles are ultra-relativistic and whose propagation direction is only barely altered by the laser field. Thus, we will keep only terms in the exponent up to order $1/p_+$ and we will neglect all quantum corrections. Since, as we will see below, the relevant proper-times $s$ scale as $1/p_+$, we can already neglect all terms containing the quantity $p_-$ in the last exponential in Eq. (\ref{D^0_i}). We point out that, even though the propagator contains an integration over all possible four-momenta $p^{\mu}$, the conservation laws and the dynamics in the processes of interest here, will automatically single out the relevant integration regions in the four-momentum and in those regions it will be $p_-\ll|\bm{p}_{\perp}|\ll p_+$. Analogously to the procedure followed in \cite{Di_Piazza_2012b,Di_Piazza_2014b}, this also allows us to neglect the field component $A_+(X)$ in Eq. (\ref{D^0_i}), which, after neglecting the terms proportional to $p_-$, always appears as a correction to the large quantity $p_+$ in the combination $p_+-eA_+(X)$ (we will come back later on this point). Finally, as we have already pointed out in Section \ref{WF}, we can also neglect the dependence of the background four-vector potential on the light-cone variable $\phi$. These considerations allow to substantially simplify the propagator $D^{(0)}(X,X'|A)$, which can be written in the form
\begin{equation}
\label{D^0_ii}
\begin{split}
D^{(0)}(X,X'|A)=&-\frac{i}{\hbar}\int_0^{\infty}ds\int\frac{d^4p}{(2\pi\hbar)^4}e^{-i[p_+(\phi-\phi')+p_-(T-T')-\bm{p}_{\perp}\cdot(\bm{x}_{\perp}-\bm{x}'_{\perp})]/\hbar}e^{i(p^2-m^2)s/\hbar}\\
&\times e^{-i(s/\hbar)\{\bm{P}^2_{\perp}+2(p_+P_T+\bm{p}_{\perp}\cdot\bm{P}_{\perp})+2e[A_-p_+-\bm{A}_{\perp}\cdot(\bm{P}_{\perp}+\bm{p}_{\perp})]+e^2\bm{A}^2_{\perp}\}}.
\end{split}
\end{equation}
This expression of the propagator precisely corresponds to the starting expression of the propagator in \cite{Di_Piazza_2012b,Di_Piazza_2014b}. Now, the possibility of neglecting the quantity $p_-$ in the last exponential brings an additional simplification as the integration in $p_-$ is trivial in Eq. (\ref{D^0_ii}). Indeed, it essentially provides a $\delta$-function $\delta(T-T'-2sp_+)$, which in turn allows to perform the proper time integration:
\begin{equation}
\label{D^0_0}
\begin{split}
D^{(0)}(X,X'|A)=&-\frac{i}{2\hbar}\int \frac{dp_+}{2\pi\hbar}e^{-ip_+(\phi-\phi')/\hbar}\frac{\theta(s_0)}{|p_+|}\int\frac{d^2\bm{p}_{\perp}}{(2\pi\hbar)^2}e^{i\bm{p}_{\perp}\cdot(\bm{x}_{\perp}-\bm{x}'_{\perp})/\hbar}e^{-i(\bm{p}_{\perp}^2+m^2)s/\hbar}\\
&\times e^{-i(s_0/\hbar)\{\bm{P}^2_{\perp}+2(p_+P_T+\bm{p}_{\perp}\cdot\bm{P}_{\perp})+2e[A_-p_+-\bm{A}_{\perp}\cdot(\bm{P}_{\perp}+\bm{p}_{\perp})]+e^2\bm{A}^2_{\perp}\}},
\end{split}
\end{equation}
where $s_0=(T-T')/2p_+$. It could have been formally more convenient to perform the integration $p_-$ later and to avoid the appearance of the $T$-dependent quantity $s_0$ in the exponential, which does not commute with the operator $P_T$. On the one hand, this does not introduce ambiguities under our approximations, as all quantum corrections in the exponentials in Eq. (\ref{D^0_0}) are neglected. On the other hand, we have ensured that the final results are unchanged if one had followed the alternative procedure and perform the integration in $p_-$ later. Thus, we can ignore this subtlety here below.

Now, the task is to disentangle the operator
\begin{equation}
O(s)=e^{-i(s/\hbar)\{\bm{P}^2_{\perp}+2(p_+P_T+\bm{p}_{\perp}\cdot\bm{P}_{\perp})+2e[A_-p_+-\bm{A}_{\perp}\cdot(\bm{P}_{\perp}+\bm{p}_{\perp})]+e^2\bm{A}^2_{\perp}\}},
\end{equation}
where $s$ is a numerical parameter. First, it is convenient to disentangle the operator $\exp[-2i(s/\hbar)(p_+P_T+\bm{p}_{\perp}\cdot\bm{P}_{\perp})]$, which corresponds to the free motion of the particle. Thus, we write
\begin{equation}
O(s)=L(s)e^{-2i(s/\hbar)(p_+P_T+\bm{p}_{\perp}\cdot\bm{P}_{\perp})},
\end{equation}
with
\begin{equation}
L(s)=e^{-i(s/\hbar)\{\bm{P}^2_{\perp}+2(p_+P_T+\bm{p}_{\perp}\cdot\bm{P}_{\perp})+2e[A_-p_+-\bm{A}_{\perp}\cdot(\bm{P}_{\perp}+\bm{p}_{\perp})]+e^2\bm{A}^2_{\perp}\}}e^{2i(s/\hbar)(p_+P_T+\bm{p}_{\perp}\cdot\bm{P}_{\perp})}.
\end{equation}
The operator $L(s)$ satisfies the differential equation
\begin{equation}
\frac{dL(s)}{ds}=-\frac{i}{\hbar}L(s)\left\{\bm{P}^2_{\perp}+2e[A_-(s)p_+-\bm{A}_{\perp}(s)\cdot(\bm{P}_{\perp}+\bm{p}_{\perp})]+e^2\bm{A}^2_{\perp}(s)]\right\},
\end{equation}
where we have introduced the short notation
\begin{align}
A_-(s)&=A_-(T-2sp_+,\bm{x}_{\perp}-2s\bm{p}_{\perp}),\\
\bm{A}_{\perp}(s)&=\bm{A}_{\perp}(T-2sp_+,\bm{x}_{\perp}-2s\bm{p}_{\perp}).
\end{align}
Since we keep only leading-order terms in $\hbar$, we can neglect the non-commutativity of the operators at different values of $s$ and write
\begin{equation}
O(s)=L(s)=e^{-(i/\hbar)\int_0^sds'\left\{\bm{P}^2_{\perp}+2e[A_-(s')p_+-\bm{A}_{\perp}(s')\cdot(\bm{P}_{\perp}+\bm{p}_{\perp})]+e^2\bm{A}^2_{\perp}(s')]\right\}}.
\end{equation}
Here, we have exploited the fact that the operator $\exp[-2i(s/\hbar)(p_+P_T+\bm{p}_{\perp}\cdot\bm{P}_{\perp})]$ in $O(s)$ acts on the unit function once the operator $O(s)$ itself is substituted back into $D^{(0)}(X,X'|A)$. Now, since $s$ scales as $1/p_+$, the only operator in the exponent of $O(s)$ of the order of unity is the one containing the quantity $p_+\int_0^s ds'A_-(s')$. According to the procedure in \cite{Di_Piazza_2012b,Di_Piazza_2014b}, it is convenient to disentangle all other operators by writing
\begin{equation}
O(s)=M(s)e^{-(i/\hbar)\int_0^sds'\left[\bm{P}^2_{\perp}-2e\bm{A}_{\perp}(s')\cdot(\bm{P}_{\perp}+\bm{p}_{\perp})+e^2\bm{A}^2_{\perp}(s')\right]},
\end{equation}
where
\begin{equation}
\begin{split}
M(s)=&e^{-(i/\hbar)\int_0^sds'\left\{\bm{P}^2_{\perp}+2e[A_-(s')p_+-\bm{A}_{\perp}(s')\cdot(\bm{P}_{\perp}+\bm{p}_{\perp})]+e^2\bm{A}^2_{\perp}(s')]\right\}}\\
&\times e^{(i/\hbar)\int_0^sds'\left[\bm{P}^2_{\perp}-2e\bm{A}_{\perp}(s')\cdot(\bm{P}_{\perp}+\bm{p}_{\perp})+e^2\bm{A}^2_{\perp}(s')\right]}.
\end{split}
\end{equation}
By applying the same procedure as in the case of the operator $L(s)$, we can write the operator $M(s)$  under the same approximations as
\begin{equation}
\begin{split}
M(s)=&e^{-(2ip_+/\hbar)\int_0^sds'A_-\big(T-2s'p_+,\bm{x}_{\perp}-2s'\bm{p}_{\perp}+2e\int_0^{s'}ds''\bm{A}_{\perp}(s'')-2s'\bm{P}_{\perp}\big)}
\end{split}
\end{equation}
and
\begin{equation}
\begin{split}
O(s)=&e^{-(2ip_+/\hbar)\int_0^sds'A_-\big(T-2s'p_+,\bm{x}_{\perp}-2s'\bm{p}_{\perp}+2e\int_0^{s'}ds''\bm{A}_{\perp}(s'')-2s'\bm{P}_{\perp}\big)}\\
&\times e^{-(i/\hbar)\int_0^sds'\left[\bm{P}^2_{\perp}-2e\bm{A}_{\perp}(s')\cdot(\bm{P}_{\perp}+\bm{p}_{\perp})+e^2\bm{A}^2_{\perp}(s')\right]},
\end{split}
\end{equation}
Now, since the argument of the second exponential already scales as $1/p_+$, we can neglect the non-commutativity of the components of the operator $\bm{P}_{\perp}$ with the corresponding coordinates and approximate the operator $O(s)$ as
\begin{equation}
\begin{split}
O(s)\approx&e^{-(2ip_+/\hbar)\int_0^sds'A_-\big(T-2s'p_+,\bm{x}_{\perp}-2s'\bm{p}_{\perp}+2e\int_0^{s'}ds''\bm{A}_{\perp}(s'')-2s'\bm{P}_{\perp}\big)}\\
&\times e^{(i/\hbar)\int_0^sds'\left[2e\bm{A}_{\perp}(s')\cdot\bm{p}_{\perp}-e^2\bm{A}^2_{\perp}(s')\right]}\\
\approx  &e^{(i/\hbar)\int_0^sds'\left[2e\bm{A}_{\perp}(s')\cdot\bm{p}_{\perp}-e^2\bm{A}^2_{\perp}(s')\right]}\\
&\times e^{-(2ip_+/\hbar)\int_0^sds'A_-\big(T-2s'p_+,\bm{x}_{\perp}-2s'\bm{p}_{\perp}+2e\int_0^{s'}ds''\bm{A}_{\perp}(s'')-2s'\bm{P}_{\perp}\big)}.
\end{split}
\end{equation}
We observe that the operator $\bm{P}_{\perp}$ cannot be neglected in the argument of the field component $A_-(\bm{X})$ because the corresponding operator in the exponential is of the order of unity. By replacing this expression of the operator $O(s)$ in Eq. (\ref{D^0_0}) and by performing the shift $\bm{p}_{\perp}\to\bm{p}_{\perp}+(\bm{x}_{\perp}-\bm{x}'_{\perp})/2s_0$ (recall that from classical electrodynamics the quantity $|\bm{x}_{\perp}-\bm{x}'_{\perp}|$ scales as $1/p_+$, see \cite{Di_Piazza_2014}), we obtain
\begin{equation}
\label{D^0_int}
\begin{split}
D^{(0)}(X,X'|A)=&-\frac{i}{2\hbar}\int \frac{dp_+}{2\pi\hbar}e^{-ip_+(\phi-\phi')/\hbar}\frac{\theta(s_0)}{|p_+|}\int\frac{d^2\bm{p}_{\perp}}{(2\pi\hbar)^2}e^{\frac{i}{\hbar}\frac{(\bm{x}_{\perp}-\bm{x}'_{\perp})^2}{4s_0}}e^{-\frac{i}{\hbar}(\bm{p}_{\perp}^2+m^2)s_0}\\
&\times e^{(i/\hbar)\int_0^{s_0}ds\{2e\bm{A}_{\perp}(s)\cdot[\bm{p}_{\perp}+(\bm{x}_{\perp}-\bm{x}'_{\perp})/2s_0]-e^2\bm{A}^2_{\perp}(s)\}}\\
&\times  e^{-(2ip_+/\hbar)\int_0^{s_0}dsA_-\big(T-2sp_+,\bm{x}_{\perp}-\frac{s}{s_0}(\bm{x}_{\perp}-\bm{x}'_{\perp})-2s\bm{p}_{\perp}+2e\int_0^sds'\bm{A}_{\perp}(s')-2s\bm{P}_{\perp}\big)},
\end{split}
\end{equation}
where from now on the meaning of the abbreviation $\bm{A}_{\perp}(s)$ is
\begin{equation}
\label{Abbr_A_perp}
\bm{A}_{\perp}(s)=\bm{A}_{\perp}\Big(T-2sp_+,\bm{x}_{\perp}-\frac{s}{s_0}(\bm{x}_{\perp}-\bm{x}'_{\perp})-2s\bm{p}_{\perp}\Big)\approx\bm{A}_{\perp}\Big(T-2sp_+,\bm{x}_{\perp}-\frac{s}{s_0}(\bm{x}_{\perp}-\bm{x}'_{\perp})\Big).
\end{equation}
The last approximation is justified because all terms containing $\bm{A}_{\perp}(s)$ in Eq. (\ref{D^0_int}) scale already as $1/p_+$ and because the integration in $\bm{p}_{\perp}$ will produce either quantum corrections (due to the factor $\exp(-is_0\bm{p}^2_{\perp}/\hbar)$, see also \cite{Di_Piazza_2012b,Di_Piazza_2014b}) or corrections scaling as $1/p_+$ (due to the factor $\exp\big[i(2e/\hbar)\bm{p}_{\perp}\cdot\int_0^{s_0}ds\bm{A}_{\perp}(s)\big]$).

At this point, the next step is to shift the variable $\bm{p}_{\perp}$ as $\bm{p}_{\perp}\to \bm{p}_{\perp}+(e/s_0)\int_0^{s_0}ds\bm{A}_{\perp}(s)$. Since after this shift the variable $\bm{p}_{\perp}$ will only produce quantum corrections, it can be neglected in the argument of the field component $A_-(\bm{X})$ and the remaining Gaussian integral in $\bm{p}_{\perp}$ can be performed analytically. The result is
\begin{equation}
\label{D^0_ip}
\begin{split}
D^{(0)}(X,X'|A)=&-\frac{1}{4\pi\hbar^2}\int \frac{dp_+}{2\pi\hbar}e^{-ip_+(\phi-\phi')/\hbar}\frac{\theta(s_0)}{|T-T'|}\exp\Big(-\frac{i}{\hbar}m^2s_0\Big)\\
&\times \exp\left\{\frac{i}{\hbar s_0}\left[\frac{\bm{x}_{\perp}-\bm{x}'_{\perp}}{2}+e\int_0^{s_0}ds\bm{A}_{\perp}(s)\right]^2-\frac{i}{\hbar}e^2\int_0^{s_0}ds\bm{A}^2_{\perp}(s)\right\}\\
&\times  e^{-(2ip_+/\hbar)\int_0^{s_0}dsA_-\big(T-2sp_+,\bm{x}_{\perp}-\frac{s}{s_0}(\bm{x}_{\perp}-\bm{x}'_{\perp})+2e\int_0^sds'\bm{A}_{\perp}(s')-2e\frac{s}{s_0}\int_0^{s_0}ds\bm{A}_{\perp}(s)-2s\bm{P}_{\perp}\big)}.
\end{split}
\end{equation}
As we have already mentioned, our aim here is the determination of the propagator within the WKB approximation. Thus, we have neglected the transverse quantum fluctuations brought about by the original presence of the term proportional to $\bm{p}_{\perp}$ in $A_-(\bm{X})$ \cite{Di_Piazza_2012b,Di_Piazza_2014b} and, apart from this aspect, the expression in Eq. (\ref{D^0_ip}) is in agreement with the corresponding one in \cite{Di_Piazza_2012b,Di_Piazza_2014b}. By expanding the last exponential operator up to terms of the order of $1/p_+$ and by updating the abbreviation $A_-(s)$ according to Eq. (\ref{Abbr_A_perp}), i.e.,
\begin{equation}
\label{A_m_s}
A_-(s)=A_-\Big(T-2sp_+,\bm{x}_{\perp}-\frac{s}{s_0}(\bm{x}_{\perp}-\bm{x}'_{\perp})\Big),
\end{equation}
we obtain
\begin{equation}
\label{D^0_1}
\begin{split}
D^{(0)}(X,X'|A)=&-\frac{1}{4\pi\hbar^2}\int \frac{dp_+}{2\pi\hbar}e^{-ip_+(\phi-\phi')/\hbar}\frac{\theta(s_0)}{|T-T'|}\exp\Big(-\frac{i}{\hbar}m^2s_0\Big)\\
&\times \exp\left\{\frac{i}{\hbar s_0}\left[\frac{\bm{x}_{\perp}-\bm{x}'_{\perp}}{2}+e\int_0^{s_0}ds\bm{A}_{\perp}(s)\right]^2-\frac{i}{\hbar}e^2\int_0^{s_0}ds\bm{A}^2_{\perp}(s)\right\}\\
&\times  \exp\left\{-\frac{4ie^2p_+}{\hbar}\int_0^{s_0}ds[\bm{\partial}_{\perp}A_-(s)]\cdot\left[\int_0^sds'\bm{A}_{\perp}(s')-\frac{s}{s_0}\int_0^{s_0}ds\bm{A}_{\perp}(s)\right]\right\}\\
&\times e^{-(2iep_+/\hbar)\int_0^{s_0}ds A_-(s)+(4iep_+/\hbar)\int_0^{s_0}ds\,s [\bm{\partial}_{\perp}A_-(s)]\cdot\bm{P}_{\perp}},
\end{split}
\end{equation}
where we have introduced the two-dimensional gradient $\bm{\partial}_{\perp}$ with respect to the whole transverse vector variable of the field to which it is applied (the transverse vector variable $\bm{x}_{\perp}-(s/s_0)(\bm{x}_{\perp}-\bm{x}'_{\perp})$ in Eq. (\ref{D^0_1})). The final task is to disentangle the last exponential in Eq. (\ref{D^0_1}). It is convenient to introduce a parameter $\rho$ and to write
\begin{equation}
\label{R}
R(\rho)=e^{-(2iep_+/\hbar)\int_0^{\rho}ds A_-(s)+(4iep_+/\hbar)\int_0^{\rho}ds\,s [\bm{\partial}_{\perp}A_-(s)]\cdot\bm{P}_{\perp}}=N(\rho)e^{(4iep_+/\hbar)\int_0^{\rho}ds\,s [\bm{\partial}_{\perp}A_-(s)]\cdot\bm{P}_{\perp}},
\end{equation}
with
\begin{equation}
N(\rho)=e^{-(2iep_+/\hbar)\int_0^{\rho}ds A_-(s)+(4iep_+/\hbar)\int_0^{\rho}ds\,s [\bm{\partial}_{\perp}A_-(s)]\cdot\bm{P}_{\perp}}e^{-(4iep_+/\hbar)\int_0^{\rho}ds\,s [\bm{\partial}_{\perp}A_-(s)]\cdot\bm{P}_{\perp}}.
\end{equation}
By differentiating with respect to the variable $\rho$ and by noticing that in general $[\bm{P}_{\perp},A_-(s)]=-i\hbar(1-s/s_0)\bm{\partial}_{\perp}A_-(s)$ (see Eq. (\ref{A_m_s})), we obtain that
\begin{equation}
N(\rho)=e^{-(2iep_+/\hbar)\int_0^{\rho}ds \left\{A_-(s)+4ep_+\left(1-\frac{s}{s_0}\right)[\bm{\partial}_{\perp}A_-(s)]\cdot\int_0^sds'\,s' [\bm{\partial}_{\perp}A_-(s')]\right\}}.
\end{equation}
In conclusion, since the operator on the right of $N(\rho)$ in Eq. (\ref{R}) acts on the unit function (once the operator $R(\rho)$ is substituted back into Eq. (\ref{D^0_1})), we obtain
\begin{equation}
\label{D^0_f}
\begin{split}
D^{(0)}(X,X'|A)=&-\frac{1}{4\pi\hbar^2}\int \frac{dp_+}{2\pi\hbar}e^{-ip_+(\phi-\phi')/\hbar}\frac{\theta(s_0)}{|T-T'|}\exp\Big[-\frac{i}{\hbar}m^2s_0-\frac{2iep_+}{\hbar}\int_0^{s_0}ds A_-(s)\Big]\\
&\times \exp\left\{\frac{i}{\hbar s_0}\left[\frac{\bm{x}_{\perp}-\bm{x}'_{\perp}}{2}+e\int_0^{s_0}ds\bm{A}_{\perp}(s)\right]^2-\frac{i}{\hbar}e^2\int_0^{s_0}ds\bm{A}^2_{\perp}(s)\right\}\\
&\times  \exp\left\{-\frac{4ie^2p_+}{\hbar}\int_0^{s_0}ds[\bm{\partial}_{\perp}A_-(s)]\cdot\left[\int_0^sds'\bm{A}_{\perp}(s')-\frac{s}{s_0}\int_0^{s_0}ds\bm{A}_{\perp}(s)\right]\right\}\\
&\times \exp\left\{-\frac{8ie^2p^2_+}{\hbar}\int_0^{s_0}ds\left(1-\frac{s}{s_0}\right)[\bm{\partial}_{\perp}A_-(s)]\cdot\int_0^sds'\,s' [\bm{\partial}_{\perp}A_-(s')]\right\}.
\end{split}
\end{equation}
As a check on this expression of the propagator, we go back to Eq. (\ref{D^0}) and we notice that the propagator $D^{(0)}(x,x'|A)$ equivalently solves the equation
\begin{equation}
\label{D^0_p}
\left[(i\hbar\partial'_{\mu}+eA_{\mu}(x'))(i\hbar\partial^{\prime,\mu}+eA^{\mu}(x))-m^2\right]D^{(0)}(x,x'|A)=\delta(x-x').
\end{equation}
This can be easily proven by writing the propagator $D^{(0)}(x,x'|A)$ in the operator form as $D^{(0)}(x,x'|A)=\langle x|D^{(0)}(A)|x'\rangle$ and by noticing that Eqs. (\ref{D^0}) and (\ref{D^0_p}) correspond to the operator equations
\begin{equation}
\begin{split}
&\left[(P_{\mu}-eA_{\mu}(X))(P^{\mu}-eA^{\mu}(X))-m^2\right]D^{(0)}(A)\\
&\qquad=D^{(0)}(A)\left[(P_{\mu}-eA_{\mu}(X))(P^{\mu}-eA^{\mu}(X))-m^2\right]=1,
\end{split}
\end{equation}
where $X^{\mu}$ and $P^{\mu}$ are four-position and the four-momentum operator, respectively (with an abuse of notation we have employed the symbol $X^{\mu}$ only in this equation to indicate the four-position operator, whereas in the rest of the paper it indicates the generic space-time point in light-cone coordinates). Now, Eq. (\ref{D^0_p}) clearly shows that the propagator $D^{(0)}(x,x'|A)$ has to fulfill the general symmetry relation $D^{(0)}(x,x'|A)=D^{(0)}(x',x|-A)$. In order to verify that indeed the propagator in Eq. (\ref{D^0_f}) satisfies this symmetry relation, it is convenient to perform the changes of variable: a) $p_+\to -p_+$, such that the quantity $s_0$ remains unchanged (after the exchange $X\leftrightarrow X'$), b) $s\to -s+s_0$ ($s'\to -s'+s_0$) in the integrations in $s$ ($s'$). 

A more explicitly symmetric form of the propagator $D^{(0)}(X,X'|A)$ can be obtained by performing the change of variable $\tau=T-2sp_+$ ($\tau'=T-2s'p_+$) in the integrals in $s$ ($s'$):
\begin{equation}
\label{D^0_fp}
\begin{split}
D^{(0)}(X,X'|A)=&-\frac{1}{4\pi\hbar^2}\frac{1}{|T-T'|}\exp\left[-\frac{i}{\hbar}e\int_{T'}^Td\tau A_-(\tau)+\frac{i}{\hbar}e\frac{\bm{x}_{\perp}-\bm{x}'_{\perp}}{T-T'}\cdot\int_{T'}^Td\tau \bm{A}_{\perp}(\tau)\right]\\
&\times\int \frac{dp_+}{2\pi\hbar}e^{-ip_+(\phi-\phi')/\hbar}\theta\left(\frac{T-T'}{2p_+}\right)\exp\left[-\frac{i}{\hbar}m^2\frac{T-T'}{2p_+}+\frac{i}{\hbar}\frac{p_+}{2}\frac{(\bm{x}_{\perp}-\bm{x}'_{\perp})^2}{T-T'}\right]\\
&\times \exp\left\langle\frac{i}{\hbar}\frac{e^2}{2p_+}\left\{\frac{1}{T-T'}\left[\int_{T'}^Td\tau\bm{A}_{\perp}(\tau)\right]^2-\int_{T'}^Td\tau\bm{A}^2_{\perp}(\tau)\right\}\right\rangle\\
&\times  \exp\Bigg\{\frac{i}{\hbar}\frac{e^2}{p_+}\int_{T'}^Td\tau [\bm{\partial}_{\perp}A_-(\tau)]\\
&\qquad\qquad\cdot\left[\frac{T-\tau}{T-T'}\int_{T'}^{\tau}d\tau' \bm{A}_{\perp}(\tau')-\frac{T'-\tau}{T-T'}\int_T^{\tau}d\tau' \bm{A}_{\perp}(\tau')\right]\Bigg\}\\
&\times \exp\Bigg\langle\frac{i}{\hbar}\frac{e^2}{2p_+}\frac{TT'}{T-T'}\\
&\qquad\qquad\times\Bigg\{\int_{T'}^Td\tau \left(1-\frac{\tau}{T'}\right)[\bm{\partial}_{\perp}A_-(\tau)]\cdot\int_T^{\tau}d\tau'\left(1-\frac{\tau'}{T}\right)[\bm{\partial}_{\perp}A_-(\tau')]\\
&\qquad\qquad-\int_{T'}^Td\tau \left(1-\frac{\tau}{T}\right)[\bm{\partial}_{\perp}A_-(\tau)]\cdot\int_{T'}^{\tau}d\tau'\left(1-\frac{\tau'}{T'}\right)[\bm{\partial}_{\perp}A_-(\tau')]\Bigg\}\Bigg\rangle,
\end{split}
\end{equation}
where
\begin{align}
\label{A_m_tau}
A_-(\tau)&=A_-\left(\tau,\bm{x}_{\perp}-\frac{T-\tau}{T-T'}(\bm{x}_{\perp}-\bm{x}'_{\perp})\right),\\
\label{A_perp_tau}
\bm{A}_{\perp}(\tau)&=\bm{A}_{\perp}\left(\tau,\bm{x}_{\perp}-\frac{T-\tau}{T-T'}(\bm{x}_{\perp}-\bm{x}'_{\perp})\right).
\end{align}
Now, let us consider the exponential function 
\begin{equation}
\mathcal{E}(\bm{X},\bm{X}')=\exp\Bigg\{\frac{i}{\hbar}\frac{e^2}{p_+}\int_{T'}^Td\tau [\bm{\partial}_{\perp}A_-(\tau)]\cdot\left[\frac{T-\tau}{T-T'}\int_{T'}^{\tau}d\tau' \bm{A}_{\perp}(\tau')-\frac{T'-\tau}{T-T'}\int_T^{\tau}d\tau' \bm{A}_{\perp}(\tau')\right]\Bigg\}
\end{equation}
in Eq. (\ref{D^0_fp}), where we notationally reminded that a dependence on the transverse variables $\bm{x}_{\perp}$ and $\bm{x}'_{\perp}$ is brought about by the components of the four-vector potential (see Eqs. (\ref{A_m_tau})-(\ref{A_perp_tau})). By integrating by parts, it can easily be shown that the quantity $\mathcal{E}(\bm{X},\bm{X}')$ can be written as
\begin{equation}
\begin{split}
\mathcal{E}(\bm{X},\bm{X}')=&\exp\Bigg\{-\frac{i}{\hbar}\frac{e^2}{p_+}\int_{T'}^Td\tau\bm{A}_{\perp}(\tau)\\
&\cdot\left[\Delta\bm{A}_{\perp}(\tau;\bm{X},\bm{X}')-\frac{1}{T-T'}\int_{T'}^Td\tau\Delta\bm{A}_{\perp}(\tau;\bm{X},\bm{X}')\right]\Bigg\},
\end{split}
\end{equation}
where
\begin{equation}
\label{Delta_A}
\begin{split}
\Delta\bm{A}_{\perp}(\tau;\bm{X},\bm{X}')=&\frac{T}{T-T'}\int_T^{\tau}d\tau'\bm{\partial}_{\perp}A_-\left(\tau',\bm{x}_{\perp}-\frac{T-\tau'}{T-T'}(\bm{x}_{\perp}-\bm{x}'_{\perp})\right)\\
&-\frac{T'}{T-T'}\int_{T'}^{\tau}d\tau'\bm{\partial}_{\perp}A_-\left(\tau',\bm{x}_{\perp}-\frac{T-\tau'}{T-T'}(\bm{x}_{\perp}-\bm{x}'_{\perp})\right).
\end{split}
\end{equation}
By manipulating in a analogous way the last exponential function in Eq. (\ref{D^0_fp}), one can finally write the propagator $D^{(0)}(X,X'|A)$ in the compact form
\begin{equation}
\label{D^0_ff}
\begin{split}
D^{(0)}(X,X'|A)=&-\frac{1}{4\pi\hbar^2}\frac{1}{|T-T'|}\exp\left[-\frac{i}{\hbar}e\int_{T'}^Td\tau A_-(\tau)+\frac{i}{\hbar}e\frac{\bm{x}_{\perp}-\bm{x}'_{\perp}}{T-T'}\cdot\int_{T'}^Td\tau \bm{A}_{\perp}(\tau)\right]\\
&\times\int \frac{dp_+}{2\pi\hbar}e^{-ip_+(\phi-\phi')/\hbar}\theta\left(\frac{T-T'}{2p_+}\right)\exp\left[-\frac{i}{\hbar}m^2\frac{T-T'}{2p_+}+\frac{i}{\hbar}\frac{p_+}{2}\frac{(\bm{x}_{\perp}-\bm{x}'_{\perp})^2}{T-T'}\right]\\
&\times \exp\left\langle\frac{i}{\hbar}\frac{e^2}{2p_+}\left\{\frac{1}{T-T'}\left[\int_{T'}^Td\tau\bm{\mathcal{A}}_{\perp}(\tau;\bm{X},\bm{X}')\right]^2\right.\right.\\
&\qquad\qquad\qquad\quad\left.\left.-\int_{T'}^Td\tau\bm{\mathcal{A}}^2_{\perp}(\tau;\bm{X},\bm{X}')\right\}\right\rangle.
\end{split}
\end{equation}
Here, we have introduced the effective transverse vector potential
\begin{equation}
\label{A_eff_prop}
\bm{\mathcal{A}}_{\perp}(\tau;\bm{X},\bm{X}')=\bm{A}_{\perp}(\tau)+\Delta\bm{A}_{\perp}(\tau;\bm{X},\bm{X}'),
\end{equation}
which exactly corresponds to the analogous effective transverse vector potential $\bm{\mathcal{A}}_{\perp}(\bm{X};T_0)$ introduced in Section \ref{WF}, in the sense that it reduces to $\bm{\mathcal{A}}^{(\text{in})}_{\perp}(\tau;\bm{x}_{\perp})$ and to $\bm{\mathcal{A}}^{(\text{out})}_{\perp}(\tau;\bm{x}'_{\perp})$ in the limits $T'\to -\infty$ and $T\to\infty$, respectively (see also the discussion below Eq. (\ref{D^0_p+})). In this respect, we want to show that, analogously to what we have observed in Section \ref{WF}, the propagator can be written in an alternative, equivalent form such that, after performing the gauge transformation $A^{\mu}(x)\to A^{\prime\mu}(x)=A^{\mu}(x)+\partial^{\mu}g(x)$ with gauge function $g(X;T_0)=-\int_{T_0}^Td\tau A_-(\tau,\bm{x}_{\perp})$, the resulting propagator in the new gauge coincides with the scalar Volkov propagator with the plane-wave transverse vector potential $\bm{A}_{\perp}(T)$ being replaced by $\bm{A}'_{\perp}(\bm{X};T_0)=\bm{\mathcal{A}}_{\perp}(\bm{X};T_0)$ (see, e.g., \cite{Brown_1964,Itzykson_b_1980}). Although this alternative expression of the propagator is formally asymmetric with respect to, e.g., the transverse variables $\bm{x}_{\perp}$ and $\bm{x}'_{\perp}$, it is instructive to sketch the proof of the above statement by reminding that under our approximations the quantity $|\bm{x}_{\perp}-\bm{x}'_{\perp}|$ scales as $1/p_+$. By starting from this observation, it is easy to show that
\begin{equation}
\begin{split}
&-e\int_{T'}^Td\tau A_-(\tau)+e\frac{\bm{x}_{\perp}-\bm{x}'_{\perp}}{T-T'}\cdot\int_{T'}^Td\tau \bm{A}_{\perp}(\tau)\\
&\qquad\approx-e\int_{T_0}^Td\tau A_-(\tau,\bm{x}_{\perp})+e\int_{T_0}^{T'}d\tau A_-(\tau,\bm{x}'_{\perp})+e\frac{\bm{x}_{\perp}-\bm{x}'_{\perp}}{T-T'}\cdot\int_{T'}^Td\tau\bm{\mathcal{A}}_{\perp}(\tau,\bm{x}_{\perp};T_0).
\end{split}
\end{equation}
Moreover, since the argument of the field-dependent exponential inside the integral in $p_+$ in Eq. (\ref{D^0_ff}) is already proportional to $1/p_+$, one can easily see from Eq. (\ref{Delta_A}) that
\begin{equation}
\begin{split}
\bm{\mathcal{A}}_{\perp}(\tau;\bm{X},\bm{X}')\approx&\bm{\mathcal{A}}_{\perp}(\tau,\bm{x}_{\perp};T_0)\\
&+\frac{T}{T-T'}\int_T^{T_0}d\tau\bm{\nabla}_{\perp}A_-(\tau',\bm{x}_{\perp})
-\frac{T'}{T-T'}\int_{T'}^{T_0}d\tau\bm{\nabla}_{\perp}A_-(\tau',\bm{x}_{\perp}).
\end{split}
\end{equation}
The proof of the above statement is complete once one observes that, since the difference $\bm{\mathcal{A}}_{\perp}(\tau;\bm{X},\bm{X}')-\bm{\mathcal{A}}_{\perp}(\tau,\bm{x}_{\perp};T_0)$ does not depend on $\tau$, then
\begin{equation}
\begin{split}
&\frac{1}{T-T'}\left[\int_{T'}^Td\tau\bm{\mathcal{A}}_{\perp}(\tau;\bm{X},\bm{X}')\right]^2-\int_{T'}^Td\tau\bm{\mathcal{A}}^2_{\perp}(\tau;\bm{X},\bm{X}')=\\
&\qquad=\frac{1}{T-T'}\left[\int_{T'}^Td\tau\bm{\mathcal{A}}_{\perp}(\tau,\bm{x}_{\perp};T_0)\right]^2-\int_{T'}^Td\tau\bm{\mathcal{A}}^2_{\perp}(\tau,\bm{x}_{\perp};T_0).
\end{split}
\end{equation}

Coming back to Eq. (\ref{D^0_ff}), we make two additional remarks. First, the argument of the exponential function outside the integral in $p_+$ coincides within our approximations with the quantity $-i(e/\hbar)(\Delta X_{\mu}/\Delta T)\int_{T'}^Td\tau A^{\mu}(\tau)$, with $\Delta X^{\mu}=(\Delta T,\Delta \bm{x}_{\perp},\Delta\phi)=(T-T',\bm{x}_{\perp}-\bm{x}'_{\perp},\phi-\phi')$, as $\Delta\phi$ scales as $1/p_+^2$. Second, the quantity $p_+$ appears in the field-dependent part of the integrand in Eq. (\ref{D^0_ff}) only in the denominator, which can be used as an \textit{a posteriori} qualitative indication that the field component $A_+(X)$ can be neglected in the propagator within our approximations (see the discussion below Eq. (\ref{D^0_i})).

Now, the integral in $p_+$ is Eq. (\ref{D^0_ff}) can be performed analytically. We first remove the $\theta$-function by splitting the integral from $-\infty$ to $+\infty$ into two integrals, one from $-\infty$ to $0$ and the other from $0$ to $+\infty$. Then, by performing the change of variable $p_+=y\Delta T/2$ in both integrals and by combining them, we obtain
\begin{equation}
\begin{split}
D^{(0)}(X,X'|A)=&-\frac{1}{16\pi^2\hbar^3}\exp\left[-\frac{i}{\hbar}e\int_{T'}^Td\tau A_-(\tau)+\frac{i}{\hbar}e\frac{\Delta\bm{x}_{\perp}}{\Delta T}\cdot\int_{T'}^Td\tau \bm{A}_{\perp}(\tau)\right]\\
&\times\int_0^{\infty} dy \exp\left\langle-\frac{i}{\hbar}\left\{\frac{\Delta X^2}{4}y+\left[m^2+\delta m^2(\bm{X},\bm{X}')\right]\frac{1}{y}\right\}\right\rangle,
\end{split}
\end{equation}
where the quantity
\begin{equation}
\begin{split}
\delta m^2(\bm{X},\bm{X}')=e^2\left\{\frac{1}{\Delta T}\int_{T'}^Td\tau\bm{\mathcal{A}}^2_{\perp}(\tau;\bm{X},\bm{X}')-\left[\frac{1}{\Delta T}\int_{T'}^Td\tau\bm{\mathcal{A}}_{\perp}(\tau;\bm{X},\bm{X}')\right]^2\right\}
\end{split}
\end{equation}
is the local generalization of the mass correction due to the laser field to the case of a focused laser field \cite{Brown_1964,Itzykson_b_1980} (see also \cite{Hebenstreit_2011b,Harvey_2012}). Finally, by performing the change of variable $y\to 1/y$, it is clear that the integrand coincides with that of the free (and of the Volkov) propagator, with the substitution $m^2\to M^2(\bm{X},\bm{X}')=m^2+\delta m^2(\bm{X},\bm{X}')\ge m^2$, such that the propagator can be written in a very compact form as \cite{Bogoliubov_b_1980}
\begin{equation}
\begin{split}
D^{(0)}(X,X'|A)=&-\frac{1}{4\pi\hbar^2}\exp\left[-\frac{i}{\hbar}e\frac{\Delta X_{\mu}}{\Delta T}\int_{T'}^Td\tau A^{\mu}(\tau)\right]\\
&\times\left[\delta(\Delta X^2)-\frac{1}{2\hbar}\frac{M(\bm{X},\bm{X}')}{\sqrt{\Delta X^2-i\eta}}\text{H}_1^{(2)}\big(M(\bm{X},\bm{X}')\sqrt{\Delta X^2-i\eta}\big)\right],
\end{split}
\end{equation}
where $\text{H}_1^{(2)}(z)$ is the Hankel function \cite{NIST_b_2010} and $\eta$ is a positive infinitesimal quantity.

Now, by introducing the reduced propagator $D^{(0)}(\bm{X},\bm{X}';p_+|A)$ such that (see Eq. (\ref{D^0_ff}))
\begin{equation}
\label{D^0_p+}
D^{(0)}(X,X'|A)=\int \frac{dp_+}{2\pi\hbar}e^{-ip_+(\phi-\phi')/\hbar}D^{(0)}(\bm{X},\bm{X}';p_+|A),
\end{equation}
it can easily be shown that it generates the in- and out-states with positive and negative energies $\Phi^{(\text{in/out})}_{\pm p}(X|A)=\exp\big[iS_{\pm p}^{(\text{in/out})}(X|A)/\hbar\big]=\exp(\mp ip_+\phi)\Phi^{(\text{in/out})}_{\pm p}(\bm{X}|A)$ of a scalar particle in agreement with the electron wave functions found in Section 2 and according to exactly the limiting procedure and substitution rules as in Eqs. (68)-(73) in \cite{Di_Piazza_2014b} (one has only to remember that here the in- and out-states are defined as those states which reduce to the free states in the remote past and future, respectively, also for the negative-energy states):
\begin{align}
\lim_{\substack{T\to\infty\\ |\bm{x}_{\perp}|\to\infty}}D^{(0)}(\bm{X},\bm{X}';p_+|A)&=-\frac{1}{4\pi\hbar^2}\frac{e^{i(\bm{p}_{\perp}^2-m^2)T/2\hbar p_+}}{T}\Phi_p^{(\text{out})*}(\bm{X}'|A) && p_+>0,\\
\lim_{\substack{T'\to-\infty\\ |\bm{x}_{\perp}'|\to\infty}}D^{(0)}(\bm{X},\bm{X}';p_+|A)&=-\frac{1}{4\pi\hbar^2}\frac{e^{i(\bm{p}_{\perp}^{\prime\,2}-m^2)|T'|/2\hbar p_+}}{|T'|}\Phi_{p'}^{(\text{in})}(\bm{X}|A) && p_+>0,\\
\lim_{\substack{T'\to\infty\\ |\bm{x}_{\perp}'|\to\infty}}D^{(0)}(\bm{X},\bm{X}';-|p_+||A)&=-\frac{1}{4\pi\hbar^2}\frac{e^{i(\bm{q}_{\perp}^2-m^2)T'/2\hbar |p_+|}}{T'}\Phi_{-q}^{(\text{out})}(\bm{X}|A) && p_+<0,\\
\lim_{\substack{T\to-\infty\\ |\bm{x}_{\perp}|\to\infty}}D^{(0)}(\bm{X},\bm{X}';-|p_+||A)&=-\frac{1}{4\pi\hbar^2}\frac{e^{i(\bm{q}_{\perp}^{\prime\,2}-m^2)|T|/2\hbar |p_+|}}{|T|}\Phi_{-q'}^{(\text{in})*}(\bm{X}'|A) && p_+<0,
\end{align}
where
\begin{align}
\label{p_q}
p^{\mu}&=p_+\left(1,\frac{\bm{x}_{\perp}}{T},1-\frac{\bm{x}^2_{\perp}}{2T^2}-\frac{m^2}{2p_+^2}\right), & q^{\mu}&=|p_+|\left(1,\frac{\bm{x}'_{\perp}}{T'},1-\frac{\bm{x}_{\perp}^{\prime\,2}}{2T^{\prime\,2}}-\frac{m^2}{2p_+^2}\right),\\
\label{pp_qp}
p^{\prime\mu}&=p_+\left(1,-\frac{\bm{x}'_{\perp}}{|T'|},1-\frac{\bm{x}_{\perp}^{\prime\,2}}{2T^{\prime\,2}}-\frac{m^2}{2p_+^2}\right), & q^{\prime\mu}&=|p_+|\left(1,-\frac{\bm{x}_{\perp}}{|T|},1-\frac{\bm{x}^2_{\perp}}{2T^2}-\frac{m^2}{2p_+^2}\right).
\end{align}

Now, by looking at Eq. (\ref{D}) for the square propagator $D(X,X'|A)$, we see that the correction in the exponential operators induced by the new spin terms scales as $1/p_+$ and it is independent of $\hbar$ (which means that it is a first-order quantum correction in $\hbar$, as the exponent in the propagator always contains the overall factor $1/\hbar$). Thus, by introducing the corresponding reduced propagator $D(\bm{X},\bm{X}';p_+|A)$ such that
\begin{equation}
D(X,X'|A)=\int \frac{dp_+}{2\pi\hbar}e^{-ip_+(\phi-\phi')/\hbar}D(\bm{X},\bm{X}';p_+|A),
\end{equation}
it is easy to show that
\begin{equation}
\label{D_f}
D(\bm{X},\bm{X}';p_+|A)=\left\{1+\frac{e}{2p_+}\gamma^+\bm{\gamma}_{\perp}\cdot[\bm{\mathcal{A}}_{\perp}(\bm{X};T_0)-\bm{\mathcal{A}}_{\perp}(\bm{X}';T_0)]\right\}D^{(0)}(\bm{X},\bm{X}';p_+|A),
\end{equation}
where it should be noticed that the difference $\bm{\mathcal{A}}_{\perp}(\bm{X};T_0)-\bm{\mathcal{A}}_{\perp}(\bm{X}';T_0)$ does not actually depend on $T_0$. Also, we point out that, analogously to the procedure followed in Section \ref{WF}, in deriving the matrix structure in Eq. (\ref{D_f}) we have retained only terms up to $1/p_+$, we have neglected corrections proportional to $1/p_+$ to the unit matrix, and we have exploited the fact that at the leading order in $1/p_+$ the second term in the curly brackets in Eq. (\ref{D_f}) is always multiplied by a matrix of the form $\bm{\gamma}\cdot\bm{e}_{\lambda}$, with $\bm{e}_{\lambda}$ being the polarization vector of a photon (almost) propagating along $\bm{n}$ (see Section \ref{WF}), and then by a free bispinor.

Now, the spinor propagator $G(X,X'|A)$ in light-cone coordinates can be obtained from Eq. (\ref{G_D}). By introducing the  reduced electron propagator $G(\bm{X},\bm{X}';p_+|A)$ such that
\begin{equation}
G(X,X'|A)=\int \frac{dp_+}{2\pi\hbar}e^{-ip_+(\phi-\phi')/\hbar}G(\bm{X},\bm{X}';p_+|A),
\end{equation}
also in the spinor case, it is
\begin{equation}
\label{G_f}
\begin{split}
G(\bm{X},\bm{X}';p_+|A)=&\left[\gamma^+\left(i\hbar\frac{\partial}{\partial T}-eA_-(\bm{X})\right)+\gamma^-p^+\right.\\
&+\bm{\gamma}_{\perp}\cdot(i\hbar\bm{\nabla}_{\perp}+e\bm{A}_{\perp}(\bm{X}))+m\bigg]D(\bm{X},\bm{X}';p_+|A).
\end{split}
\end{equation}
By exploiting the arguments below Eq. (\ref{A_eff_prop}) and by noticing that
\begin{equation}
\label{comm}
\begin{split}
&\left[\gamma^+\left(i\hbar\frac{\partial}{\partial T}-eA_-(\bm{X})\right)+\gamma^-p^++\bm{\gamma}_{\perp}\cdot(i\hbar\bm{\nabla}_{\perp}+e\bm{A}_{\perp}(\bm{X}))+m\right]\\
&\quad\times\exp\left[-i\frac{e}{\hbar}\int_{T_0}^Td\tau A_-(\tau,\bm{x}_{\perp})+i\frac{e}{\hbar}\int_{T_0}^{T'}d\tau A_-(\tau,\bm{x}'_{\perp})\right]\\
&=\exp\left[-i\frac{e}{\hbar}\int_{T_0}^Td\tau A_-(\tau,\bm{x}_{\perp})+i\frac{e}{\hbar}\int_{T_0}^{T'}d\tau A_-(\tau,\bm{x}'_{\perp})\right]\\
&\quad\times\left[\gamma^+i\hbar\frac{\partial}{\partial T}+\gamma^-p^++\bm{\gamma}_{\perp}\cdot(i\hbar\bm{\nabla}_{\perp}+e\bm{\mathcal{A}}_{\perp}(\bm{X};T_0))+m\right],
\end{split}
\end{equation}
it can be seen that also in the spinor case the propagator can be written alternatively in an equivalent form such that, after performing the gauge transformation mentioned below Eq. (\ref{A_eff_prop}), the resulting propagator in the new gauge coincides with the spinor Volkov propagator with the plane-wave transverse vector potential $\bm{A}_{\perp}(T)$ being replaced by $\bm{A}'_{\perp}(\bm{X};T_0)=\bm{\mathcal{A}}_{\perp}(\bm{X};T_0)$. In order to show this statement, it is also important to observe that within our approximations and after the commutation in Eq. (\ref{comm}) has been carried out, the action of the operator $i\hbar\bm{\nabla}_{\perp}$ in Eq. (\ref{G_f}) on the background four-vector potential can be ignored. This formal equivalence indicates that the propagator $G(X,X'|A)$ can also be written in the compact form
\begin{equation}
G(X,X'|A)=\int \frac{d^4p}{(2\pi\hbar)^4}E_p(X;T_0|A)\frac{\hat{p}+m}{p^2-m^2+i\eta}\bar{E}_p(X';T_0|A),
\end{equation}
like in the plane-wave case \cite{Ritus_1985}, where $\hat{p}=\gamma^{\mu}p_{\mu}$, where in general $\bar{E}_p(X;T_0|A)=\gamma^0E^{\dag}_p(X;T_0|A)\gamma^0$, and where the dependence on $T_0$ finally compensates out.

Finally, it can be easily shown that the analogous limiting procedures as in Eqs. (74)-(77) in \cite{Di_Piazza_2014b} allow to obtain the in- and out-electron states in agreement with Eqs. (\ref{in-out_pm_i})-(\ref{in-out_pm_f})): 
\begin{align}
\label{V_T_infty}
\lim_{\substack{T\to\infty\\ |\bm{x}_{\perp}|\to\infty}}G(\bm{X}_{\perp},\bm{X}'_{\perp};p_+|A)&=-\frac{1}{4\pi\hbar^2}\frac{e^{i(\bm{p}_{\perp}^2-m^2)T/2\hbar p_+}}{T}\sum_{\sigma=1}^2\sqrt{2\varepsilon}\,u_{p,\sigma}\bar{\psi}_{p,\sigma}^{(\text{out})}(\bm{X}'|A) && p_+>0,\\
\lim_{\substack{T'\to-\infty\\ |\bm{x}'_{\perp}|\to\infty}}G(\bm{X}_{\perp},\bm{X}'_{\perp};p_+|A)&=-\frac{1}{4\pi\hbar^2}\frac{e^{i(\bm{p}_{\perp}^{\prime\,2}-m^2)|T'|/2\hbar p_+}}{|T'|}\sum_{\sigma=1}^2 \sqrt{2\varepsilon}\,\psi_{p',\sigma}^{(\text{in})}(\bm{X}|A)\bar{u}_{p',\sigma} && p_+>0,\\
\lim_{\substack{T'\to\infty\\ |\bm{x}'_{\perp}|\to\infty}}G(\bm{X}_{\perp},\bm{X}'_{\perp};-|p_+||A)&=\frac{1}{4\pi\hbar^2}\frac{e^{i(\bm{q}_{\perp}^2-m^2)T'/2\hbar |p_+|}}{T'}\sum_{\sigma=1}^2\sqrt{2\varepsilon}\,\psi_{-q,-\sigma}^{(\text{out})}(\bm{X}|A)\bar{u}_{-q,-\sigma} && p_+<0,\\
\label{V_Tp_infty}
\lim_{\substack{T\to-\infty\\ |\bm{x}_{\perp}|\to\infty}}G(\bm{X}_{\perp},\bm{X}'_{\perp};-|p_+||A)&=\frac{1}{4\pi\hbar^2}\frac{e^{i(\bm{q}_{\perp}^{\prime\,2}-m^2)|T|/2\hbar |p_+|}}{|T|}\sum_{\sigma=1}^2 \sqrt{2\varepsilon}\,u_{-q',-\sigma}\bar{\psi}_{-q',-\sigma}^{(\text{in})}(\bm{X}'|A) && p_+<0,
\end{align}
where $\bar{\psi}(x)=\psi^{\dag}(x)\gamma^0$ for a generic bispinor $\psi(x)$ and where, consistently with the results in Section \ref{WF}, the first-order terms in $1/p_+$ in the pre-exponent proportional to the unit matrix have been neglected (the
four-momenta $p^{\mu}$, $p^{\prime,\mu}$, $q^{\mu}$, and $q^{\prime,\mu}$
are given in Eqs. (\ref{p_q})-(\ref{pp_qp})). Obviously, the in-and out-states in Eqs. (\ref{V_T_infty})-(\ref{V_Tp_infty}) differ from those in Eqs.  (\ref{in-out_pm_i})-(\ref{in-out_pm_f}) only by the $\phi$-dependent phase factors $\exp(-ip_+\phi)$ and $\exp(ip_+\phi)$ for the positive- and negative-energy states, respectively, i.e., $\psi_{\pm p,\pm\sigma}^{(\text{in/out})}(X|A)=\exp(\mp ip_+\phi)\psi_{\pm p,\pm\sigma}^{(\text{in/out})}(\bm{X}|A)$. Also, it should be noticed that in Eqs. (\ref{V_T_infty})-(\ref{V_Tp_infty}) the quantity $\varepsilon=\sqrt{m^2+\bm{p}^2}$ is always a positive quantity and the appearance of the quantity $\sqrt{2\varepsilon}$ instead that of $2|\epsilon|$ as in \cite{Di_Piazza_2014b} is only due to a difference in the normalization of the constant bispinors. Finally, it is also worth observing that, since in deriving the propagators we have neglected all quantum corrections in the exponents, the limiting procedure cannot reproduce the correcting terms proportional to the unit matrix in the pre-exponent in Eq. (\ref{psi_0}), as those arising, e.g., from the Van Vleck determinant. However, as we have mentioned in Section \ref{WF}, such terms can be neglected in the calculation of the probabilities of strong-field QED processes. 

%
%
\section{Conclusions and outlook}
\label{Concl}
In conclusion, in the present paper we have continued and extended the investigation started in \cite{Di_Piazza_2014}, and provided analytical tools for studying strong-field QED processes in the presence of tightly focused laser beams. As in \cite{Di_Piazza_2014}, we have focused on the regime of interaction where the asymptotic energy of the electron is the largest dynamical energy scale in the problem. First, we have presented an equivalent but simpler expression of the wave functions found in \cite{Di_Piazza_2014}. We have also generalized the expression of the wave functions, including the possibility that the electron also has a non-vanishing asymptotic transverse momentum. This is particularly important for the study of first-order strong-field QED processes, where the two leptons (e.g., initial and final electron or final electron and positron) do not propagate exactly along the same direction. Moreover, we have shown that the wave functions found here have essentially the same structure of the Volkov states and that the former can be practically obtained from the latter by means of a simple substitution of the transverse vector potential of the laser field.

Moreover, we have now constructed the corresponding scalar and spinor propagators under the same approximations and we have verified by means of an appropriate limiting procedure that the determined propagator is compatible with the corresponding wave functions. We have seen that the structure of the propagator is also essentially the same as the Volkov propagator, with an effective transverse vector potential also containing derivatives of the focused laser field with respect to the transverse variables. This has allowed for the introduction of a local laser-induced mass correction also in the case of a focused laser field.

In an upcoming publication we will apply the wave functions found here and in \cite{Di_Piazza_2014} to investigate the first-order strong-field QED processes (nonlinear Compton scattering and nonlinear Breit-Wheeler pair production) including the tight focusing of the background laser field.

%
%
\section*{Acknowledgments}

This work has been partially carried out during the KITP program ``Frontiers of Intense Laser Physics'' and fruitful discussions there with G. V. Dunne, K. Z. Hatsagortsyan, and C. Schubert are gratefully acknowledged. During that period this research was supported in part by the National Science Foundation under Grant No. NSF PHY11-25915. I am also grateful to E. Yakaboylu and to A. I. Milstein for useful discussions on the WKB propagator and on gauge transformations, and on the quasiclassical approximation, respectively.

\bibliography{Bibliography/Books,Bibliography/Reviews,Bibliography/Papers_Radiation,Bibliography/Papers_RR,Bibliography/Papers_PP_and_Cascades,Bibliography/Papers_VPE,Bibliography/Papers_Various}

\end{document}